\documentclass{aa}  

\usepackage{graphicx}
\usepackage{txfonts}
\usepackage{comment}
\usepackage{amsmath}
\usepackage{float}

\usepackage{xcolor}

\begin{document}

   \title{Neural blind deconvolution to reconstruct high-resolution ground-based solar observations}

   \author{C. Schirninger
          \inst{1}
          \and
          R. Jarolim\inst{2}
          \and
          A. M. Veronig\inst{1,}\inst{3}
          \and
          M. Rempel\inst{2}
          \and
          F. W{\"o}ger\inst{4}
          }

   \institute{Institute of Physics, University of Graz,
              Universitätsplatz 5, 8010 Graz, Austria
        \and
            High Altitude Observatory, NSF National Center for Atmospheric Research, 3080 Center Green Dr, Boulder, USA  
        \and
            Kanzelhöhe Observatory for Solar and Environmental Research, University of Graz, 
            Treffen am Ossiacher See, Austria
        \and
            National Solar Observatory, 3665 Discovery Drive, Boulder, CO 80303, USA
            }

  \abstract
   {Ground-based solar observations enable unprecedented spatial, spectral, and temporal resolution of the lower solar atmosphere, yet Earth’s turbulent atmosphere imposes significant limitations, requiring advanced post-facto image reconstruction. State-of-the-art reconstruction methods are based on restoring a burst of short exposure frames to a single observation. Limitations of these techniques arise due to the sparse information about the atmospheric point spread function (PSF) that degrade the observations and consequently the quality of reconstructions. Developing new reconstruction methods is essential for providing high quality data products for the study of the lower solar atmosphere on the smallest scales.}
   {We develop a novel image reconstruction method to achieve unprecedented spatial resolution from short exposure image bursts. This can provide high-quality reconstructions and therefore advance the study of the smallest spatial scales from the solar photosphere to the chromosphere.}
   {In this study, we present a novel approach for high-resolution solar image reconstruction based on physics-informed neural networks. In the training process, the neural network maps coordinate points $(x,y)$ directly to their corresponding intensity values $o(x,y)$ while simultaneously updating the PSF parameters. The method convolves the "true" object from the neural network with the estimated PSFs and optimizes the network by minimizing the loss between the synthesized and real short-exposure image burst. This approach enables the simultaneous estimation of both the degrading PSF and the real high-resolution intensity distribution.}
   {We demonstrate the method on synthetic intensity data derived from a radiative MHD simulation, where we apply PSF convolution and noise to obtain a realistic synthetic input data set, similar to observational short-exposure observations. Quantitative comparisons using image quality metrics, histograms, and power spectral analysis confirm that the model can reliably reconstruct the original image from the stack of synthetic short-exposure frames. Finally, we apply our method to high-resolution observations from GREGOR and DKIST, and compare to state-of-the-art speckle reconstructions and multi-frame blind deconvolutions. Our results demonstrate the ability to reconstruct small-scale solar features that exceed the reconstruction performance of state-of-the-art reconstruction methods. With this approach we lay the foundation for future spatially varying PSFs.}
   {}

   \keywords{Image processing --
                image reconstruction --
                atmospheric seeing --
                high resolution --
                telescopes --
                solar physics 
               }

   \maketitle
%

\section{Introduction}

Large-aperture solar telescopes provide high resolution observations, enabling detailed studies on very small scales in the different atmospheric layers of the Sun, from the solar photosphere to the lower solar corona. Such high-quality observations are essential for studying the Sun's dynamic processes and magnetic activity. Turbulent plasma motions within the convection zone and strong magnetic fields are key drivers of energetic eruptive phenomena, including solar flares and coronal mass ejections \citep{priest2002, wiegelmann2014}.

However, observations from large-aperture ground-based telescopes are affected by Earth's turbulent atmosphere. While the wavefront from a distant point source can be considered as flat outside Earth's atmosphere, atmospheric turbulence introduces phase errors that lead to image degradation and blurring in the observational data \citep{mckechnie1992, quirrenbach2006}.  State-of-the-art high-resolution telescopes are equipped with adaptive optics (AO) systems, which aim to minimize the phase errors in the wavefront and thus reduce the distortion in the observations \citep{rimmele2011}. Nevertheless, due to turbulence in multiple layers of the Earth's atmosphere, these corrections are challenging, which limits the spatial and spectral resolution of the observations \citep{rimmele2008}. Equipped with four-meter mirrors, the Daniel K. Inouye Solar Telescope (DKIST; \citealp{rimmele2020}) and the future European Solar Telescope (EST; \citealp{noda2022}) offer unprecedented spatial, spectral, and temporal resolution for studying the solar atmosphere. However, they also require sophisticated image reconstruction methods to achieve their full resolution potential. Post-facto image reconstructions provide the only option to account for the remaining atmospheric degradations and to achieve the diffraction-limited resolution of large-aperture telescopes. \citep{loefdahl2002}. 

\begin{figure*}[h!tbp]
   \sidecaption
   \centering
   \includegraphics[scale=0.065]{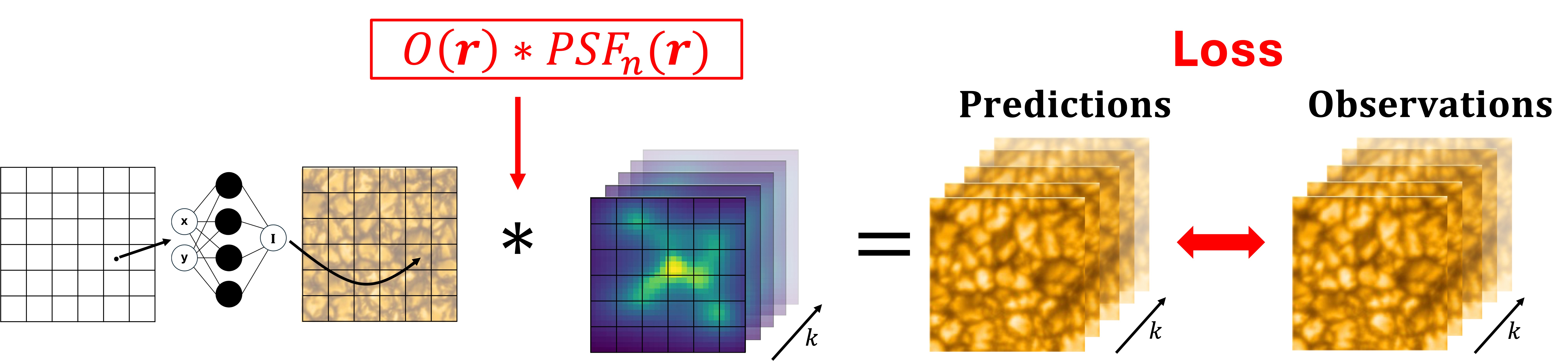}
   \caption{Overview of the NeuralBD deconvlution pipeline. The neural networks maps coordinate points to pixel intensity values to estimate the real object intensity distribution. The convolution is calculated between the estimated real object and the predicted $n$ PSFs which are parameterized as learnable values on a grid. The optimization is performed between the predicted image burst and the original image burst from the telescope.}
\label{fig:nbd-pipeline}
\end{figure*}

State-of-the-art reconstruction methods such as speckle reconstruction, multi-frame blind deconvolution (MFBD) and multi-object multi frame blind deconvolution (MOMFBD) are based on the assumption that the real object is distorted by the convolution with an unknown point spread function (PSF) and an additional noise term \citep{knox1974, boer1992, loefdahl2003, schulz1993, vanNoort2005}. Short exposure observations preserve information in the higher frequency domain, quasi "freezing" the state of the turbulent atmosphere, to retain the diffraction-limited information \citep{labeyrie1970}. For a high cadence image burst it can be assumed that the solar features remain unchanged while the atmospheric degradations change on a much shorter time scale. Consequently, using a burst of short exposure observations improves the quality of the reconstruction \citep{Woeger2008}. The image formation of the burst can be described as the convolution of the real object $o(r)$ with the PSFs $p_k(r)$ and the addition of a noise terms $n_k(r)$, which is primarily attributed to photon noise
\begin{equation}
\label{eq:image}
    i_{k}(r) = o(r) \ast p_{k}(r) + n_{k}(r)\,,
\end{equation}
where $r=(x,y)$ are the coordinates in the image space and $k$ indexes individual image frames, $k = 1 .... K$, with $K$ representing the total number of frames in a burst. The PSFs are composed of the telescope PSF and the PSFs from the turbulent atmosphere. 

For MFBD and MOMFBD this problem is ill-posed for a single image, since neither the real object nor the PSFs of the turbulent atmosphere are known for the deconvolution process. In order to solve this problem to estimate $o(r)$, a stack of short-exposure images is used and is described by \cite{vanNoort2005}. Here, the image formation (Eq.\,\ref{eq:image}) is expressed in the Fourier domain as
\begin{equation}
\label{eq:image_fourier}
    I_{k}(u) = O(u) \cdot P_{k}(u) + N_{k}(u),
\end{equation}
where $P_{k}(u)$ is the optical transfer function (OTF), i.e. the Fourier transform of the PSF. In MFBD, the PSF is initialized through a wavefront in the pupil plane. The phase is constructed with a finite set of of basis functions $\text{B}_m(\xi)$ and wavefront coefficients $\alpha_{k,m}$ following
\begin{equation}
    \varphi_k(\xi) = \sum\limits_{m=0}^M \alpha_{k,m} \text{B}_m(\xi).
\end{equation}
Using this phase representation, the OTF can be written as
\begin{equation}
P_k(u) = \mathcal{F} \left\{ \left| \mathcal{F}^{-1} \left[A(\xi) \,\exp\!\left( i \, \varphi_k(\xi) \right)\right] \right|^2 \right\},
\end{equation}
with $A(\xi)$ corresponding to the pupil function. Finally, solving Eq.\,\ref{eq:image_fourier} corresponds to a joint least-squares optimization over the object and the wavefront coefficients
\begin{equation}
\label{eq:mfbd_optim}
\mathcal{L}=\sum_{k=1}^{K}\sum_{u}\left|I_k(u)-\hat{O}(u) \,\hat{P}_k(u)\right|^2,
\end{equation}
where $\hat{P}_k(f)$ and $\hat{O}(f)$ correspond to estimated quantities. Under the assumption of additive gaussian noise, the estimation of true object intensities, $\hat{O}(u)$, can be written as
\begin{equation}
\hat{O}(u)=H(u)\frac{\sum_{k}I_k(u) \,P_k^*(u, \hat{a}_k)}{\sum_{k} \left|P_k(u, \hat{a}_k)\right|^{2}},
\end{equation}
reducing the least-squares optimization to the $\hat{P}_k(u)$ only. Here, $^*$ denotes complex conjugations and $H(u)$ corresponds to a low-pass filter from e.g. \cite{lofdahl1994}.

In contrast to MFBD and MOMFBD, speckle reconstruction follows a statistical approach in which the Fourier amplitudes and phases are estimated separately. The Fourier phases are recovered using a generalization of the Knox–Thompson method, the speckle masking technique, which evaluates the bispectrum (triple correlations in Fourier space) to obtain robust phase estimates of the true object \citep{lohmann1983}. The Fourier amplitudes are determined following the method of \cite{labeyrie1970}, which requires knowledge of the speckle transfer function (STF).
To determine the STF, the spectral ratio is computed and compared with atmospheric turbulence models to estimate the Fried parameter ($r_0$), which is a measure of the strength of the atmospheric seeing \citep{vonderluehe1984}. Once the STF is known, the true object’s spatial power spectrum can be reconstructed and together with the recovered phases an estimate of the true object intensity distribution can be obtained \citep{Woeger2008}.

Limitations of state-of-the-art methods can arise from reconstruction artifacts as well as their high computational cost. The methods work under isoplanatic conditions (non-field-dependent PSF), meaning that the wavefront distortions are constant across the image. Consequently, only small patches can be reconstructed, which limits the effectiveness when a large field of view (FOV) needs to be restored. Mosaicing these patches can enable large FOV reconstructions, but can also introduce artifacts in the restorations. Furthermore, the reconstruction methods currently in use are based in the Fourier domain. Transforming back and forth between the Fourier and image domain can introduce artifacts, which may further degrade the reconstruction quality. Additionally, estimating the true PSF is challenging. MFBD makes assumptions using a set of basis functions and wavefront coefficients to estimate the PSF, which can make it difficult to fully recover the true PSF. Errors in PSF estimation propagate into the reconstructed image, causing artifacts or loss of resolution \citep{lofdahl1994, vanNoort2005, Lofdahl2007}.

To overcome these limitations, \cite{ramos2018} first applied convolutional neural networks to reconstruct a burst of degraded observations. In a follow-up paper, \cite{ramos2023} used deep neural networks to accelerate the deconvolution problem, employing the same approach as the MOMFBD method. This involves using a set of basis functions to simultaneously estimate the PSF and the real object. Following the MOMFBD methodology, \cite{ramos2024} employed a neural emulator network with a spatially variant PSF to mitigate the effects of anisoplanatism. \cite{schirninger2025} applied generative adversarial networks (GANs; \citealp{goodfellow2020}) to reconstruct a burst of observation to a single high quality, high-resolution observation.

The proposed NeuralBD method is a new reconstruction method where we solve the image equation given in Eq.\,\ref{eq:image}. The method is based on Physics-informed neural networks (PINNs; \citealp{raissi2019, karniadakis2021}), which have already been shown to successfully incorporate physical models with noisy observational data in solar physics \citep{jarolim2023, jarolim2024, baso2025, jarolim2025}. Our NeuralBD model uses a neural network to encode input coordinates ($x, y$) to pixel intensity values $o(x,y)$. The training process involves estimating the PSF, computing the convolution in the image domain and minimizing the residuals between the estimated and original burst of observation. Using a PINN enables a smooth representation of the image while requiring relatively low memory, as no training dataset is needed. Instead, the image equation (Eq.\,\ref{eq:image}) is solved iteratively, allowing simultaneous estimation of both the PSFs and the true intensity distribution of the object. This represents a novel approach compared to state-of-the-art reconstruction methods, as the problem is solved directly in the image domain rather than in the Fourier domain. This lays the foundation to extend this method to spatially varying PSFs allowing reconstruction over larger FOV.

In this study, we present an image reconstruction method based on the simultaneous estimation of the distorting PSF and the real object. In our approach, the PSF does not consider any constraints except pixel size, by iteratively updating the PSF parameters in the training step. NeuralBD is a solver-based method rather than a function approximator like classical deep learning models. Consequently, it is more computational expensive, but provides solutions independent of any reference data set. This makes the approach applicable to any high-resolution broadband imaging instrument and offers the possibility to find more effective solutions of the PSF and therefore, the real object reconstruction. We demonstrate that our NeuralBD method can estimate the real object and the PSF simultaneously. To test the model and to quantify its performance, we apply it to simulation data and real observations.

\section{Method}
\label{sec:method}
The aim of our NeuralBD method is to estimate the real object intensity distribution, by solving the inverse problem as described in Eq.\ref{eq:image}. The neural network serves as a smooth function approximation of the real object $o(x,y)$ where coordinate points ($x, y$) are mapped to intensity values $o(x,y)$. By iteratively updating the real object and the PSFs we calculate the convolution to match the observed image burst $i_{k}(x,y)$. This is contrary to MFBD, which performs the least-squares optimization in the Fourier domain (Eq.\,\ref{eq:mfbd_optim}) and only depends on the estimation of the OTFs.

We approximate the continuous object intensity distribution by learning a mapping $(x,y) \mapsto o(x,y)$. Our pixel coordinates are first centered around zero and then normalized by dividing by 1024. We sample the normalized coordinate points and apply a random perturbation within each grid cell, rather than evaluating the network only at pixel centers. This perturbed sampling avoids restricting the model to fixed pixel centers and instead enables the network to represent a continuous intensity distribution, which corresponds to a stochastic sub-pixel sampling of the underlying image domain.

The second step in our NeuralBD pipeline is to estimate the true PSFs for the deconvolution process. In contrast to conventional approaches of constraining the PSFs with a set of basis functions and wavefront coefficients, we fully model the PSF as a set of $N\times N$ pixels free parameters from an arbitrary distribution. The initialization of the PSFs is performed using a centered normal distribution with a mean of zero and a standard deviation of five pixels ($\mu=0, \sigma=5$). However, the initialization can be chosen arbitrarily, e.g., by sampling from a random distribution or also using the Karhunen-Lo{\'e}ve polynomials with the wavefront coefficients. For the deconvolution process, the number of PSFs corresponds to the number of frames in the burst ($n$). The only constraint we set for the PSF is that each PSF must sum to one and must be positive, following 
\begin{equation}
\label{eq:PSF_discrete}
    \sum_{x'} \sum_{y'} \text{PSF}_k(x',y') = 1, \quad \text{PSF}_k(x',y') > 0 \quad \forall x',y',
\end{equation}
Based on the selected number of frames, we generate $K$ corresponding PSFs to perform the convolution. The 2D PSF parameters are optimized by initializing them as learnable parameters within the model and scale them, where the output of the PSFs corresponds to $\exp({\log\_{\text{PSF}_{k}}})$, where $k$ corresponds to the index of the PSF. 

Third, we calculate the convolution of the estimated object intensity distribution $o(x,y)$ using the estimated PSFs in the image space. We approximate the continuous convolution
\begin{equation}
\tilde{i}_k(x,y) = \iint \tilde{o}(x+\xi, y+\eta)\, \tilde{p}_k(\xi,\eta)\, d\xi\, d\eta
\end{equation}
by a discrete quadrature over the perturbed sampling locations,
\begin{equation}
\label{eq:conv_area}
\tilde{i}_k(x,y)=\sum_{x'} \sum_{y'}\tilde{o}(x+x',y+y') \,\tilde{p}_k(x',y') \,a(x',y'),
\end{equation}
where $\tilde{i}_k$ denotes the estimated short-exposure image burst, $\tilde{o}$ the true object intensity distribution and $\tilde{p}_k$ the corresponding PSFs estimated by the NeuralBD model. 
The term $a(x',y')$ indicates the local area elements associated with each perturbed sampling point and corresponds to the quadrature weights approximating the differential surface element $dA = d\xi\,d\eta$. These weights are computed from the distances between adjacent perturbed sampling locations and ensure conservation of PSF unity in the discretized convolution. After calculating the convolution, we employ a mean squared error (MSE) loss computed between the synthesized burst and the original burst of observation for the optimization of the neural network and the PSF parameters. The optimization is carried out through an iterative process during which both the real object intensity distribution and the PSF parameters are progressively updated.

The NeuralBD image model is implemented as a fully connected neural network consisting of eight hidden layers with 512 neurons each. We utilize a sine activation function on each layer, inspired by the sinusoidal representation networks (SIREN; \citealp{sitzmann2020}). Before we feed the coordinate points into the first layer, we employ a Gaussian positional encoding. Consequently, we encode the input coordinates into Fourier features of $sin$ and $cos$ functions. Here, we choose 128 random frequencies from a normal distribution scaled between 0 and 4 ($f_i \sim \mathcal{N}(0, 4)$), which we found to give the best results. The resulting encoding is given by
\begin{equation}
    \begin{split}
    \nu_{\text{encoded}} = [\sin(f_{x,0}\cdot x), \cos(f_{x,0}\cdot x), \sin(f_{x,1}\cdot x), \\ \cos(f_{x,1}\cdot x),\,...\,, \sin(f_{y,127}\cdot y), \cos(f_{y,127}\cdot y)], 
    \end{split}
\end{equation}
with $f_{i,j}$ corresponding to the $j$th sampled frequency of coordinate $i$. Incorporating Gaussian positional encoding, allows for better selection of high frequency terms producing sharper reconstructions, while the comparison with a classical SIREN network resulted in less sharp reconstruction and slower convergence. The frequencies of the encoding are decisive for the success of the method, as it controls the ability to encode high frequency features \citep{tancik2020}. Frequencies that are too high may introduce artifacts in the reconstruction, whereas frequencies that are too low fail to capture fine structural details. Additionally, the Fourier feature regularization acts as implicit noise damping due to the models bias towards lower frequency representations. Consequently, NeuralBD does not require a noise filter. The input layer then takes the encoded coordinate points and the neural network maps them to the intensity values $o(x,y)$. For output activation, we found that an exponential function with base 10 gives the best results. To accelerate the reconstruction process, we first pre-train the image model $o(x,y)$ for 300 epochs using the original frame with the lowest root-mean-square (RMS) error as reference. This initialization allows the model to start from a reasonable estimate of the object, rather than learning the image from scratch, and consequently focus directly on the reconstruction task. Note, that in principle reconstruction from scratch is also possible, but the reconstruction time would be significantly increased, while performance is comparable.

To validate the performance of our NeuralBD method, we consider three image reconstruction methods for comparison, a classical deconvolution technique and two state-of-the-art reconstruction methods for high-resolution solar observations. In this study, the Richardson–Lucy (RL) deconvolution serves as the baseline reconstruction method \citep{richardson1972}. We apply the RL deconvolution to the simulation data, where we use the known ground truth PSFs as input for the reconstruction. The number of iterations is set to 1000, which we found empirically to provide the best results in terms of sharpness and stability. A lower number of iterations leads to less sharp reconstructions. As a second comparison, we employ a multi-frame blind deconvolution approach using the torchmfbd method of \cite{ARamos2025}. This method is used for the comparison of both, the simulations data and the high-resolution observations obtained from GREGOR and DKIST. In addition, we use speckle reconstructions for the comparison of high-resolution observational data. For GREGOR, the speckle reconstructions are obtained using the KISIP software package from \cite{woeger2008_kisip}, while for DKIST they are carried out using the method described in \cite{beard2020}. To evaluate the power spectral density (PSD) of each observation including the convolved frames from the simulation data, the original frames from the real observations, the NeuralBD reconstructions, and the high-quality reference observations, we use and adapt the azimuthal PSD implementation from \cite{ramos2024}. The PSD is calculated as a function of spatial frequency, which depends on the spatial sampling of the observations.

\begin{table}[h!tbp]
\caption{Comparison of the quality metrics between the baseline (Richardson-Lucy) the torchmfbd and our NeuralBD model against the MURaM simulation.}             
\label{table:quality-metrics}      
\centering                          
\begin{tabular}{c | c c c}        
\hline\hline                 
& MSE & SSIM & PSNR \\       
\hline \\
Richardson Lucy & 0.0067 & 0.49 & 21.7 \\
& & &  \\
torchmfbd & 0.0038 & 0.57 & 24.22 \\
& & &  \\
NeuralBD & 0.0004 & 0.96 & 34.41 \\ 
\hline 
\end{tabular}
\tablefoot{The comparison is made using the mean squared error (MSE), the structural similarity index measure (SSIM) and the peak signal-to-noise ratio (PSNR).}
\end{table}

\section{Data}
In this study, we evaluate the performance of our NeuralBD model using three data sets. The first data set is based on synthetic data generated using the MURaM simulation code \citep{rempel2016}. The second consists of real high-resolution observations from the 1.5 m GREGOR telescope \citep{schmidt2012}. The third data set corresponds to observations from the 4 m DKIST telescope \citep{rimmele2020}.

\subsection{MURaM simulation}
\label{sect:muram}
To confirm that our NeuralBD method can find the correct solution for the reconstructed observations we test the model on synthetic data derived from the MURaM radiative magnetohydrodynamics (MHD) simulation by \cite{rempel2012}. The data corresponds to synthetic integrated intensity images from a MURaM simulation with a resolution of 4096$\times$4096 pixels and 12\,km grid spacing. We use binned and downsampled data with a resolution of 1024$\times$1024 pixels and a spatial sampling of 96\,km pixels$^{-1}$. It shows sunspot with umbra, penumbra and quiet Sun regions surrounding the sunspot, providing a realistic test case for the application to solar observations. 

In order to test the deconvolution performance of the NeuralBD model, we degrade the high quality simulation first. We synthesize $n$ PSFs based on the MFBD and MOMFBD approach, using Karhunen-Lo{\'e}ve basis functions and a set of wavefront coefficients to model the aberrations, since this is the current state-of-the-art for reconstruction methods \citep{vanNoort2005}. These PSFs are then convolved with the intensity images derived from the simulation data. To create the synthesized PSFs we use the implementation of \cite{ramos2024}. The wavefronts are calculated with
\begin{equation}
    \varphi(\xi) = \sum\limits_{m=0}^M \alpha_l \text{KL}_m(\xi),
\end{equation}
where $\alpha_l$ corresponds to the wavefront coefficients generated randomly from a uniform distribution in the interval [$-2, 2$] and KL$_m$ to the Karhunen-Lo{\'e}ve basis functions using 44 modes. The pupil function is then given by
\begin{equation}
    P(\xi)=A(\xi)e^{i\varphi(\xi)},
\end{equation}
where $A(\xi)$ corresponds to the amplitude of the pupil which we consider as constant ($A(\xi)=1$). Finally, we compute the PSF by taking the inverse Fourier transform of the pupil function
\begin{equation}
    \text{PSF}(r)=|\mathcal{F}^{-1}(P(\xi))|^2,
\end{equation}
where $r$ corresponds to the spatial coordinates ($x,y$). 

To generate $n$ distinct PSFs, we compute $n$ unique wavefronts by applying $n$ different wavefront coefficients. Each wavefront is then used to derive its corresponding PSF. We chose a pixel size of 29$\times$29 pixels. Note that a larger PSF can be sampled, however we found that $29\times29$ is sufficiently large for our applications and a larger PSF would require additional computing resources.

In the reconstruction process, we first normalize the simulation intensity in an interval between 0 and 1. Second, we crop a small region from the full FOV with a size of $24\times24$ Mm. Lastly, we convolve the simulation with 50 distinct synthesized PSFs as described above to obtain an image burst of 50 short exposure observations. To generate a more realistic degraded burst of observations, Gaussian noise is added during the convolution process, using the mean and standard deviation derived from the simulation data.

\begin{figure}[h!tbp]
   \centering
   \includegraphics[scale=0.070]{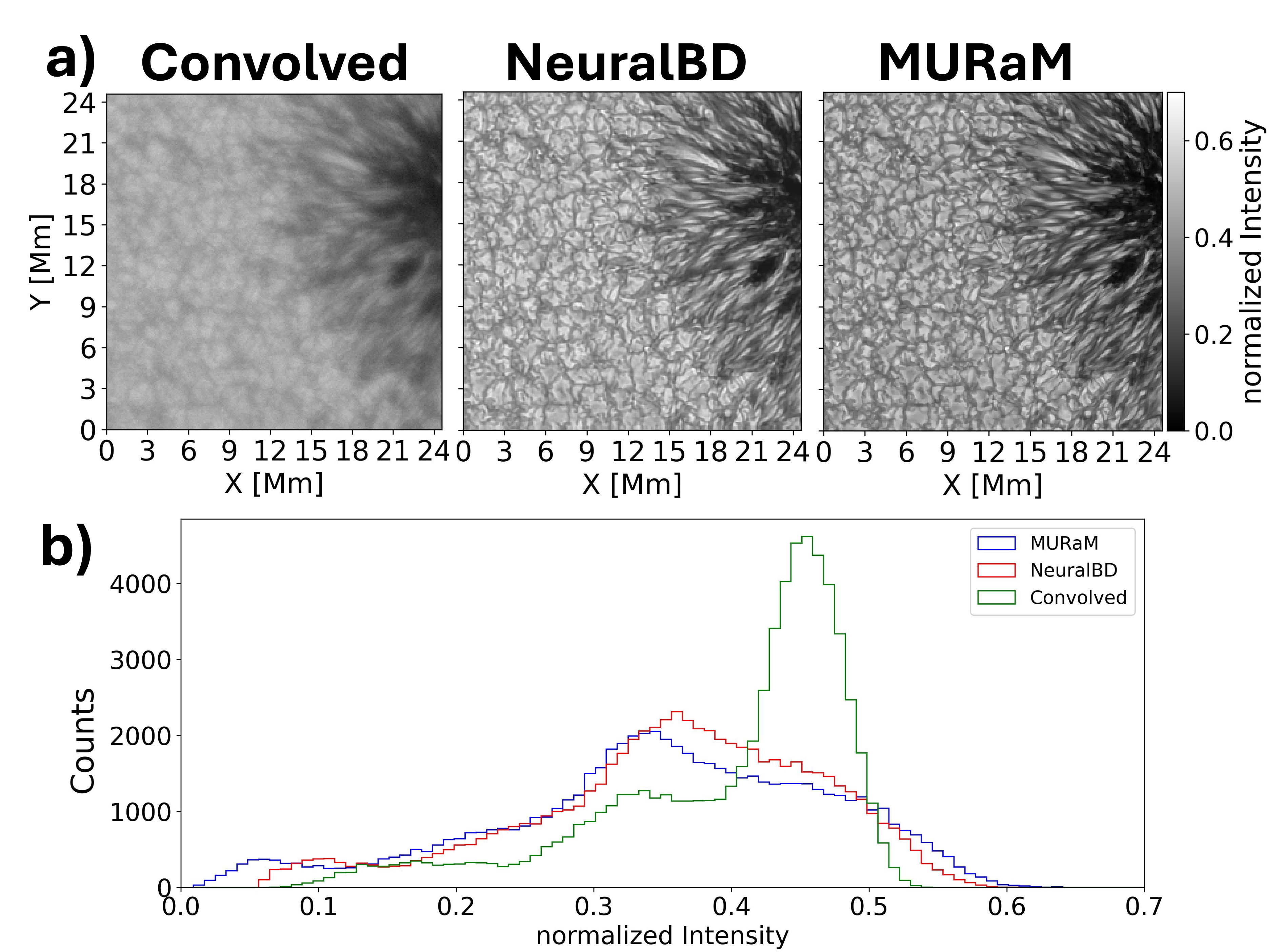}
   \caption{a) Comparison of a single frame of the burst from the degraded simulation (left), the NeuralBD reconstruction (center) and the real simulation (right). b) shows the histogram of the normalized intensity for the convolved frame (green), the NeuralBD reconstruction (red) and the real simulation (blue).}
\label{fig:muram_conv-rec}
\end{figure}

\subsection{GREGOR observations}
\label{sec:gregor_obs}
With the application to observational data we demonstrate the applicability of our NeuralBD model to real observations. Here we consider high-resolution observations from the GREGOR High-Resolution Fast Imager (HiFI; \citealp{denker2023}). The instrument provides high spatial and temporal imaging in six wavelength bands with formation heights in the solar photosphere to the chromosphere. The FOV covers approximately $75''$ corresponding to a spatial resolution of $0.028''\text{pixel}^{-1}$. 

The observations we use were obtained on 2 June 2022 at 09:50:15 UT and consists of an image burst of 100 frames in the blue continuum (450.6 nm) and G-Band (430.7 nm). The observation target is a small sunspot with umbra, penumbra and surrounding granulation, similar to the MURaM simulation. The image quality for this observation is good (good seeing), but still shows noise and perturbations from Earth's atmosphere, requiring post-facto image reconstruction.

We pre-process the original image burst before feeding the observations into the neural network. First, we crop a smaller region from the full FOV with a size of $20''\times20''$. All frames are then aligned to the first frame, which serves as reference. The alignment is performed by optimizing the spatial shift derived from cross-correlation analysis. This step accounts for potential spatial shifts in the observations caused by atmospheric turbulence. Second, we crop a smaller region with a size of  $7''\times7''$ from the aligned observations. Third, we normalize the image burst to an interval between 0 and 1. Finally, the burst is sorted by RMS contrast, with the observations that have the best RMS being prioritized. For the reconstruction, the best 50 frames are selected. The reconstructions for the two wavelength bands are performed separately, even though the observations are acquired simultaneously. Due to a small difference in the light path, the observations of the two wavelength bands are rotated and shifted. To align them and therefore simultaneously reconstruct both observations simultaneously would require more extensive preprocessing.  

\begin{figure*}[h!tbp]
   \centering
   \includegraphics[scale=0.11]{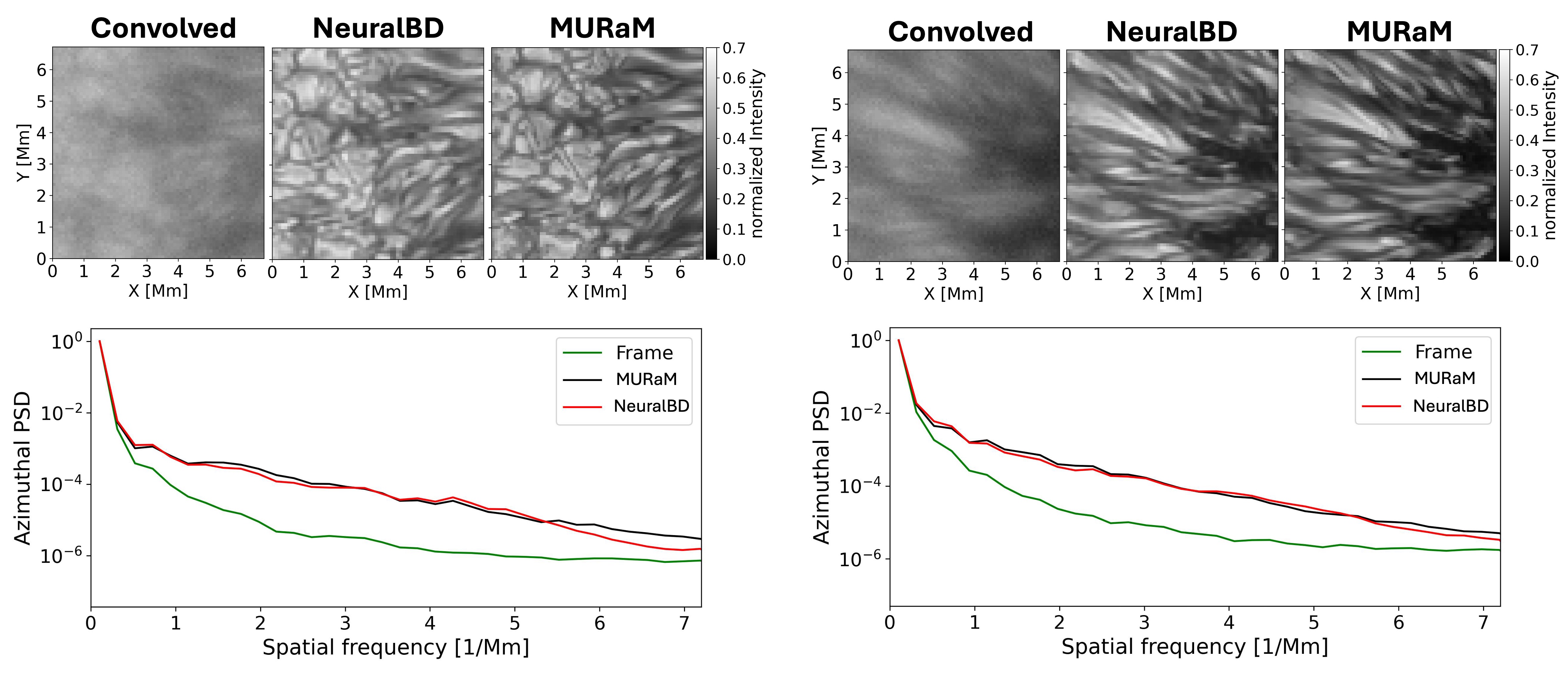}
   \caption{Comparison of the degraded convolved frame (left) the NeuralBD reconstruction (center) and the real simulation (right) for quiet Sun and penumbra and umbra with penumbra. The bottom panel shows the corresponding power spectra density (PSD) for the convolved frame (green), the NeuralBD reconstruction (red) and the real simulation (black).}
\label{fig:muram_recon_psd}
\end{figure*}

\subsection{DKIST observations}
As a second high-resolution observation application we test our NeuralBD method on observations from the 4\,m DKIST telescope \citep{rimmele2020}. Specifically, we use data from the Visible Broadband Imager (VBI; \citealp{woger2021}) observing the solar photosphere and chromosphere with high spatial and temporal resolution. Our test data consist of an observation of the instruments blue arm centered on the blue continuum at 450.3\,nm. The  observed field of view is $45'' \times 45''$  recorded with $4\,\text{k} \times4\,\text{k}$ pixels, which corresponds to a pixel resolution of $0.011''\text{pixel}^{-1}$.

The observation is part of a test data set from DKIST and consists of 80 short-exposure frames. The observation target is a big sunspot with smaller spots and granulation. Although the observation quality is high due to good seeing conditions, applying post-facto image reconstruction can further enhance the data by compensating for residual atmospheric effects due to the telescope's large aperture. 

We pre-process the DKIST observation by cropping out a smaller FOV of $12''\times12''$. The next step involves aligning the 80 short exposure frames using cross-correlation analysis, with the first frame acting as the reference. After that we crop out a region with the size of $5.5''\times5.5''$ and normalize the intensity values based on the global intensity maximum of the burst for each frame to an interval between 0 and 1. Following the same approach as for GREGOR, we sort the image burst by RMS contrast, listing the observations with the highest RMS contrast first, to prioritize the highest quality frames.

\section{Results}
The training of the NeuralBD method involves estimating simultaneously both the real object intensity distribution using the image model and the PSFs, as described in Sect.\,\ref{sec:method}. For training, we set the number of epochs to 8000 using a batch size of 1024. The total reconstruction time corresponds to approximately $7$\,h on a NVIDIA A100 GPU.

To assess the performance of our NeuralBD method, we conduct a quantitative evaluation using simulation data and provide a qualitative comparison with observational data. For the application to real observations we compare to state-of-the-art multi-frame blind deconvolutions and speckle reconstructions which are currently implemented in the instrument pipeline of the telescopes.

\subsection{Quantitative comparison - MURaM}
To evaluate the reconstruction performance, we test our NeuralBD model on synthetic data from a MURaM simulation. The burst of degraded simulations, as described in Sect.\,\ref{sect:muram} is used as input to the NeuralBD pipeline to reconstruct the real simulation.

We evaluate the performance of the model using three quality metrics. The Mean Squared Error (MSE) serves as a measure of distortion quality. The perceptual quality of the reconstruction is evaluated using the Structural Similarity Index Measure (SSIM; \citealp{wang2004}) and the Peak Signal-to-Noise Ratio (PSNR; \citealp{fardo2016}). In order to perform a comparative evaluation, we compare our NeuralBD reconstructions against the baseline reconstruction method using the Richardson-Lucy deconvolution and the torchmfbd method as described in Sect.\,\ref{sec:method}. Table\,\ref{table:quality-metrics} summarizes all three quality metrics of our NeuralBD reconstruction with the comparison to the baseline reconstruction method and torchmfbd. NeuralBD outperforms both, the baseline and torchmfbd in all three quality metrics. For MSE, lower values indicate a closer resemblance to the ground truth simulation, whereas higher values of SSIM and PSNR correspond to better perceptual quality. Note that the Richardson-Lucy deconvolution uses the known PSF as input. In contrast, NeuralBD and torchmfbd estimates the true object intensity distribution as well as the PSFs simultaneously, however, NeuralBD is able to outperform both methods.

In Fig.\,\ref{fig:muram_conv-rec}a) we show a single frame of the degraded burst of the simulation (left), our NeuralBD reconstruction (center) and the real simulation (right). The observation with a FOV of 24$\times$24 Mm shows part of the sunspot, the penumbra and quiet Sun regions. Our NeuralBD reconstruction demonstrates a clear improvement in both large-scale and small-scale features when comparing the convolved frame with NeuralBD and MURaM. The granulation pattern in the quiet Sun region, as well as the penumbra shows more details compared to a single frame from the degraded burst. Additionally, the noise present in the convolved frame is mitigated by the NeuralBD reconstruction, due to inherent spatial smoothness of the reconstructions \citep{jarolim2025}. Furthermore, the contrast of the NeuralBD reconstruction shows a higher similarity to the real MURaM simulation. This is also confirmed in Fig.\,\ref{fig:muram_conv-rec}b) which sows the intensity distribution of the single convolved frame, the simulation and our NeuralBD reconstruction.

Figure\,\ref{fig:muram_recon_psd} shows smaller crops of the single frame of the burst, the NeuralBD reconstruction and the real simulation of Fig\,\ref{fig:muram_conv-rec}. Additionally, we visualize the PSD for each of the crops. The lower panels shows the power spectra for the single frame of the burst (green), the NeuralBD reconstruction (red), and the ground truth simulation (black) as a function of spatial frequency. The power spectra of the NeuralBD reconstruction closely match those of the real simulation for both crops. Despite the strong overall agreement, small deviations can be seen in the spatial frequency range from 1 to 3 Mm$^{-1}$, corresponding to the reconstruction performance of middle sized structures.

\begin{figure}[h!tbp]
   \centering
   \includegraphics[scale=0.12]{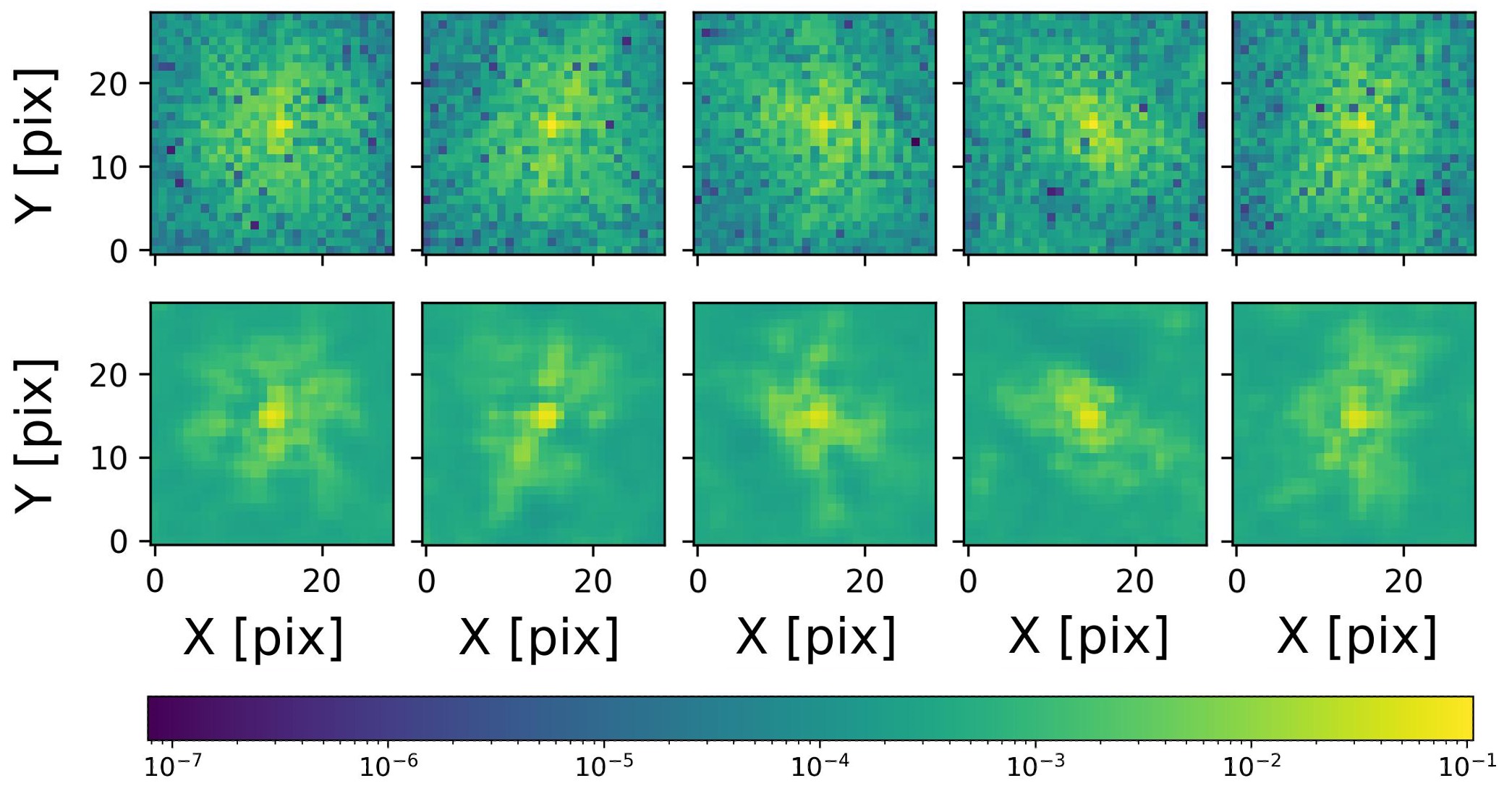}
   \caption{Comparison of five PSFs initialized to degrade the synthetic MURaM simulation (top panel) and the PSFs estimation by our NeuralBD model (bottom panel).}
\label{fig:muram_psfs}
\end{figure}

\begin{figure*}[h!tbp]
   \centering
   \includegraphics[scale=0.085]{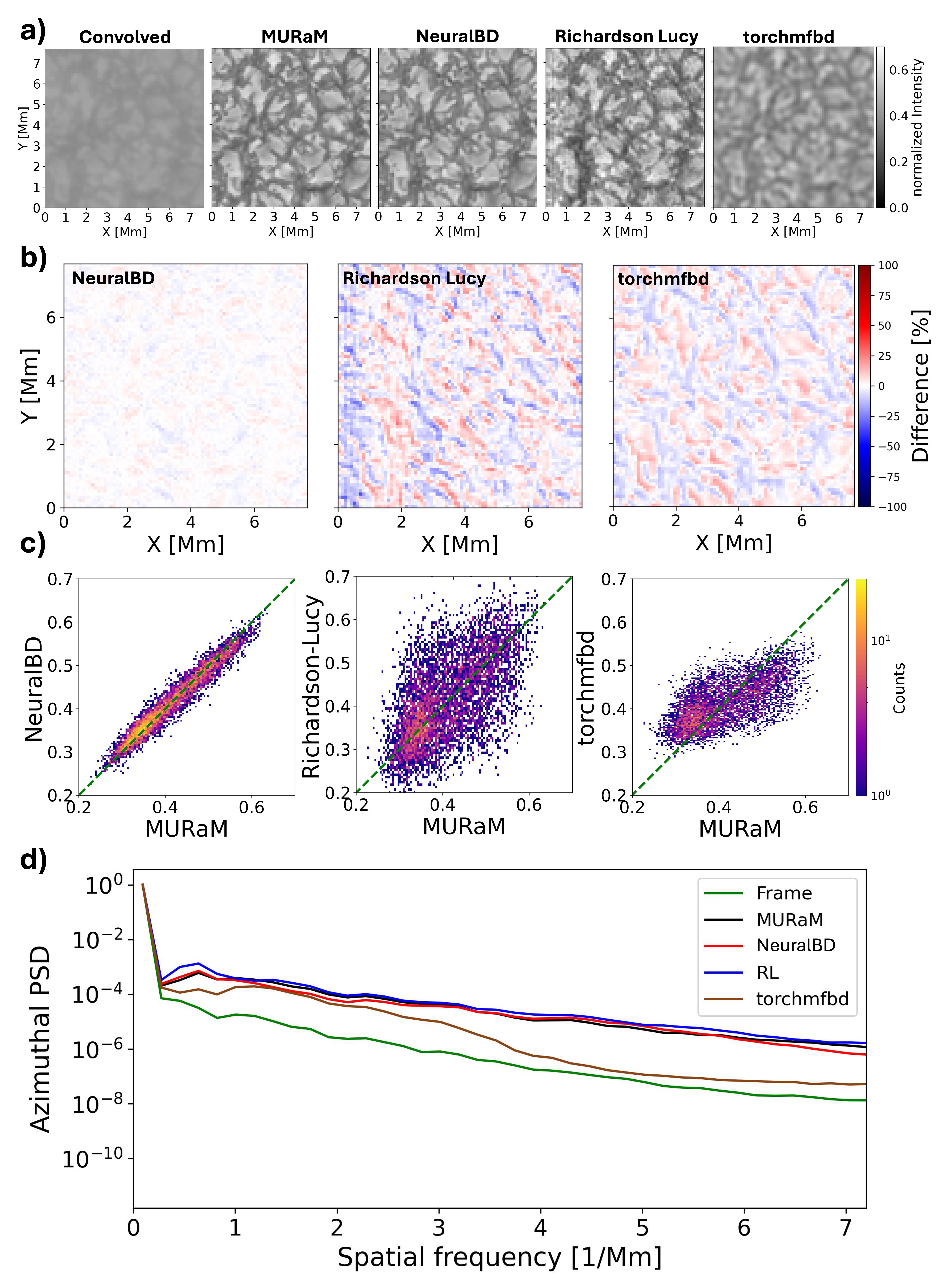}
   \caption{Comparison between the baseline method (Richardson-Lucy), the torchmfbd method and our NeuralBD reconstruction for MURaM simulation data. In panel a) we show a visual comparison of a single convolved frame, the MURaM simulation, the NeuralBD reconstruction and the Richardson-Lucy deconvolution and torchmfbd. Panel b) shows the difference maps of our NeuralBD method, the baseline method and torchmfbd. Panel c) shows the 2d histograms comparing NeuralBD with MURaM (left) and Richardson-Lucy with MURaM (center) and torchmfbd with MURaM (right). In panel d) the power spectral density (PSD) for all five observations in a) are shown.}
\label{fig:muram_rl}
\end{figure*}

\begin{figure*}[h!tbp]
   \centering
   \includegraphics[scale=0.09]{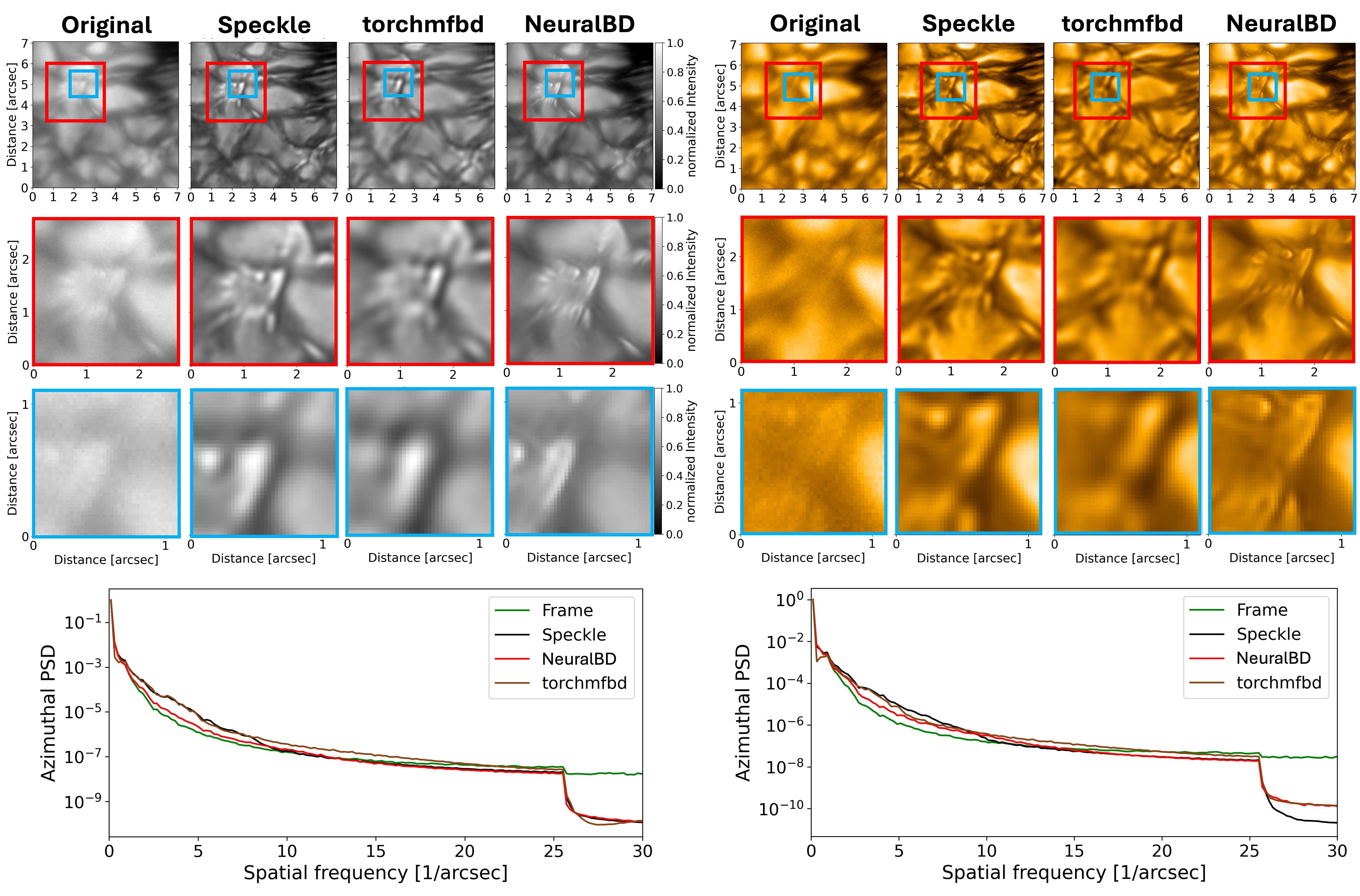}
   \caption{Comparison of the performance of the NeuralBD reconstuction method with the speckle reconstruction and torchmfbd for g-band at 430.7 nm (left) and blue continuum at 450.6 nm (right). a) Single frame from the original burst (first column), the speckle reconstruction (second column), the torchmfbd reconstruction (third column) and the NeuralBD reconstruction (fourth column), for the GREGOR observation on June 2, 2022. The red and blue rectangles indicate the regions shown as cropped views in the second and third rows, respectively. b) Corresponding azimuthal power spectra for both channels are shown. The single frame of the original burst (green), the speckle reconstruction (black), the torchmfbd reconstruction (brown) and the NeuralBD reconstruction (red).}
\label{fig:gregor_con_recon}
\end{figure*}
Since the ground truth PSFs are available for the reconstruction in Fig.\,\ref{fig:muram_conv-rec}, we can directly compare them with those estimated by our NeuralBD method. Figure\,\ref{fig:muram_psfs} illustrates the effectiveness of our model in recovering accurate PSFs. The top row shows an example of the first five ground truth PSFs used to degrade the simulation before training, while the bottom row shows the corresponding PSFs estimated by the NeuralBD model. The estimated PSFs closely resemble the ground truth in terms of their overall structure and features, demonstrating the model’s ability to produce realistic solutions. There are small differences on the smaller scales, as the NeuralBD PSFs look more smooth. The cause for this is the intialization of the PSF as a Gaussian distribution. NeuralBD can also be initialized using the Karhunen-Lo{\'e}ve polynomials as basis functions and estimate the wavefront coefficients to obtain the exact solution. However, the PSF is always learned in the image domain, without any constraints independent of predefined wavefront coefficients. As starting initialization, we use a gaussian distribution, but the model has the freedom to converge to any solution. The results shown in Fig.\,\ref{fig:muram_psfs} demonstrate that our NeuralBD method can effectively handle the additional degree of freedom to estimate both, the real object intensity distribution and the PSFs by using just a Gaussian distribution as initialization.

\begin{figure*}[h!tbp]
   \sidecaption
   \centering
   \includegraphics[scale=0.045]{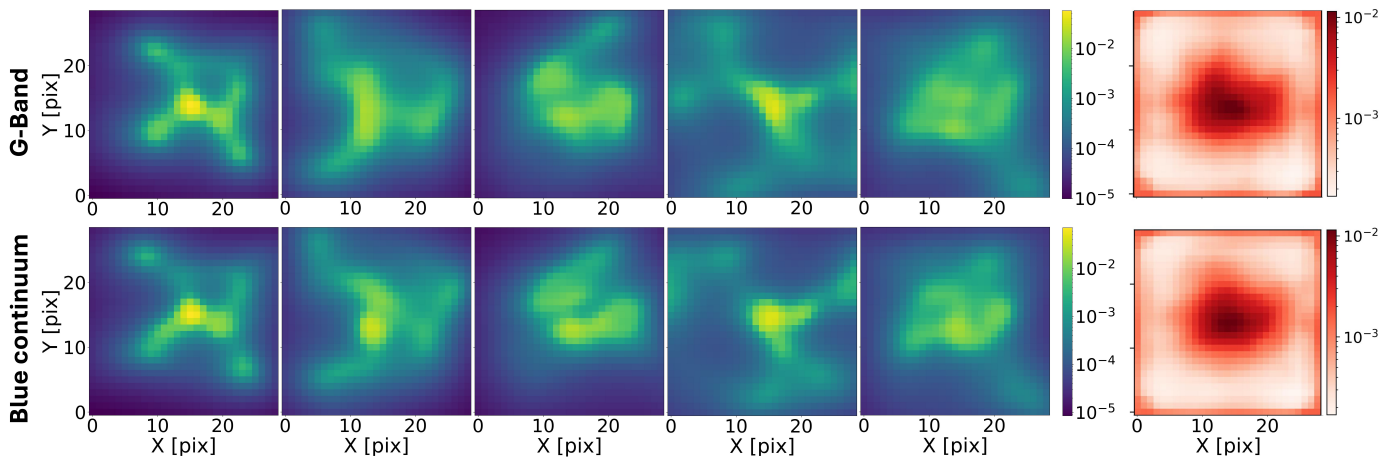}
   \caption{Comparison of the the first five PSFs at 430.7 nm (G-Band) in the first row and 450.6 nm (Blue continuum) in the second row, estimated by our NeuralBD model. The last column corresponds to the mean PSF calculated over the 50 PSFs from the NeuralBD model.}
\label{fig:gregor-psfs}
\end{figure*}

In Fig.\,\ref{fig:muram_rl} we compare the reconstruction performance of our NeuralBD method compared to a standard reconstruction method using the Richardson-Lucy deconvolution and the torchmfbd method. Figure\,\ref{fig:muram_rl}a) shows a clear improvement of the reconstructed observations for, NeuralBD, Richardson-Lucy and torchmfbd compared to the single convolved frame. However, comparing NeuralBD, Richardson-Lucy and torchmfbd to the ground-truth MURaM simulation, NeuralBD shows a much closer resemblance to the simulation. Richardson-Lucy has much higher contrast and shows small scale artifacts in the reconstruction. On the other hand, torchmfbd shows a less sharp reconstruction compared to NeuralBD. This is also confirmed by the difference maps in Fig.\,\ref{fig:muram_rl}b). The difference between NeuralBD and the ground-truth MURaM simulation (left) is much lower as compared to the difference between the Richardson-Lucy deconvolution and the MURaM simulation (center) and the difference between torchmfbd and \nobreak{MURaM} (right). The color bar represents the relative difference in percent. This is also confirmed by Table\,\ref{table:quality-metrics} where NeuralBD outperforms both Richardson-Lucy and torchmfbd in all three quality metrics, including both, distortion and perceptual metrics. The 2d histogram in Fig.\,\ref{fig:muram_rl}c) also demonstrates the close resemblance of the NeuralBD reconstruction to the MURaM simulation compared to the Richardson-Lucy deconvolution and the torchmfbd method. It is important to note that the baseline method uses a single ground-truth PSF for deconvolution. In contrast, our NeuralBD approach estimates the PSFs and the reconstructed observation simultaneously during training, which makes the task considerably more challenging yet yields to better results. Additionally, the torchmfbd method requires the information of the size of the telescope aperture and the wavelength for the reconstruction. Both parameters are not available since we use a MURaM simulation as ground truth. Consequently, we tested different configurations of telescope aperture and wavelength and we found an aperture of 5\,m and a wavelength of 450\,nm give the best results. An exhaustive hyperparameter search could further improve the reconstructions. To complete the evaluation, Fig.\,\ref{fig:muram_rl}d) shows the PSD for the convolved frame, the MURaM simulation, the NeuralBD reconstruction, the Richardson-Lucy deconvolution and the torchmfbd method. Here, Richardson-Lucy deconvolution and NeuralBD show close similarity with the simulation, outperforming the degraded frame. The reduced power of the torchmfbd method can be caused by the missing information of the telescope aperture and the wavelength. Note that Richardson-Lucy deconvolution uses only one frame for reconstruction, whereas NeuralBD and torchmfbd use 50 frames and NeuralBD the estimated PSFs by the neural network. 

\subsection{Application to GREGOR data}
\label{sec:gregor_app}
To demonstrate the reconstruction performance of our NeuralBD model on real high-resolution observations we apply it to data from the 1.5\,m GREGOR telescope and compare our NeuralBD reconstructions to state-of-the-art speckle reconstructions \citep{kuckein2017} and multi-frame blind deconvolutions. Fig.\,\ref{fig:gregor_con_recon}a) shows observations from June 2, 2022 in G-Band at 430.7 nm and blue continuum at 450.6 nm, respectively. The first column for each wavelength corresponds to a single frame with the best RMS contrast of the original image burst, the second column to the speckle reconstruction, the third column to the torchmfbd reconstruction and the fourth column to our NeuralBD reconstruction. In the second and third rows, we show smaller crops marked by the red and blue boxes. For comparison, we shift the NeuralBD reconstruction to match the mean value of the speckle reconstruction. This adjustment is necessary because the speckle reconstruction incorporates an atmospheric model with the speckle transfer function to estimate true intensities, whereas the NeuralBD reconstruction does not. However, such a correction can be applied afterwards. All three, NeuralBD, speckle and the torchmfbd reconstructions show a clear improvement in sharpness compared to the original single frame. On the largest scales, NeuralBD, speckle and the torchmfbd reconstructions demonstrate similar image quality. However, when zooming in on smaller scales (red box), NeuralBD reveals finer, more detailed structures within the bright points for both wavelength bands. At sub-arcsec scales (blue box), NeuralBD reveals features that are visible in the original frame, but remain unresolved and appear blurred in the speckle and the torchmfbd reconstruction.

In Fig.\,\ref{fig:gregor_con_recon}b) we show the corresponding PSD evaluations for the single frame, the speckle reconstruction, the torchmfbd reconstruction and the NeuralBD reconstruction. All three, speckle, the torchmfbd and NeuralBD reconstructions outperform single-frame data with torchmfbd and speckle showing similar power, slightly enhanced compared to NeuralBD. The blue continuum reconstruction performs better than the G-band reconstruction and even yields more power than the speckle reconstruction in the spatial frequency range from 10 to 14 arcsec$^{-1}$. However, within the 4 to 9 arcsec$^{-1}$ spatial frequency range, the speckle reconstruction shows slightly higher power in both filter bands. The feature around 25 arcsec$^{-1}$ in both wavelength bands where the power suddenly drops corresponds to the diffraction limit of the telescope. 

\begin{figure*}[h!tbp]
   \centering
   \includegraphics[scale=0.14]{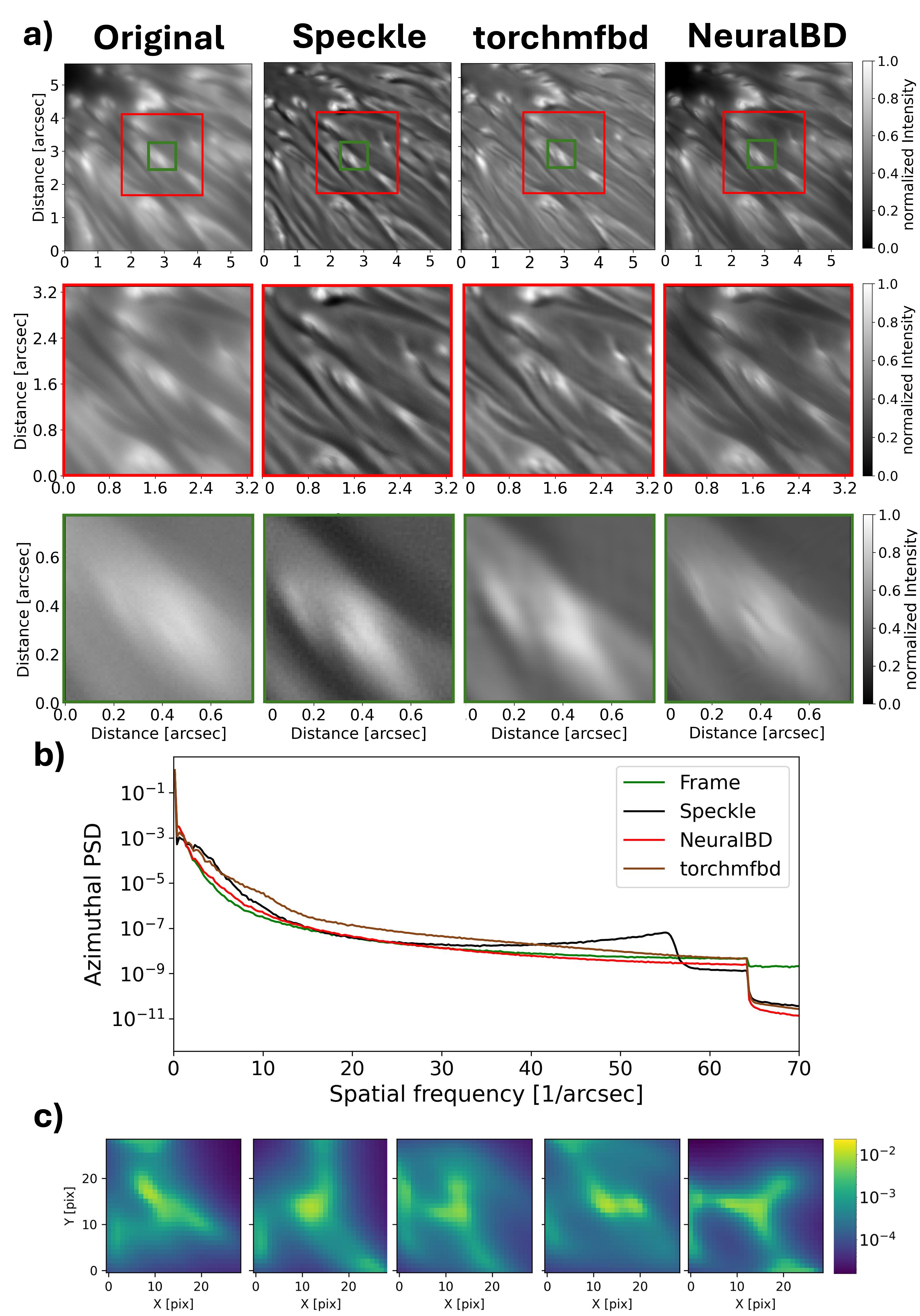}
   \caption{Comparison of the reconstruction methods with NeuralBD, torchmfbd and speckle. a) Single frame of the original burst (first column), the speckle reconstruction (second column), the torchmfbd (third column) and the NeuralBD reconstruction (fourth column) is shown. The red and green boxes show zoomed in regions on smaller spatial scales. b) The azimuthal power spectra is shown where the green line corresponds to the single frame of the burst, black to the speckle reconstruction, brown to the torchmfbd reconstruction and red to the NeuralBD reconstruction. c) Five example point spread functions estimated by the NeuralBD method.}
\label{fig:dkist}
\end{figure*}

Fig.\,\ref{fig:gregor-psfs} shows five representative examples of the first five frames from the 50 PSFs estimated by our NeuralBD model. The first row shows the PSFs estimated for the G-band, and the second row shows the PSFs for the blue continuum. The overall structure of the PSFs are very similar in both bands, suggesting that despite the large degree of freedom of the modeled PSFs a unique solution exists. Additionally, the PSFs show a significant amount of jittering which cannot be modeled simply using wavefront coefficients. This is also confirmed by the last column which shows the average PSF calculated over all estimated 50 PSFs from the NeuralBD model. However, there are small differences visible at smaller scales, which can be caused by the difference in the two filters that observe the same scene. Solving Eq.\,\ref{eq:image} in the image domain also allows for the construction of arbitrary PSFs, including spatial shifts, and is essential for extending the approach to spatially varying PSFs.

\subsection{Application to DKIST data}
As a second high-resolution observation application, we test our NeuralBD method on data from the VBI instrument on DKIST and compare our NeuralBD reconstruction with state-of-the-art multi-frame blind deconvolutions and speckle reconstructions. Fig.\,\ref{fig:dkist}a) shows the comparison of a single frame of the original burst (first column) the speckle reconstruction (second column), the torchmfbd reconstruction (third column) and our NeuralBD reconstruction (fourth column). The $5.5''\times5.5''$ cutout of the observation shows umbra and penumbra regions. In the 2nd and 3rd row we show smaller crops, zooming in on penumbral structures up to sub-arcsecond scales. Comparing the single frame of the burst with the best RMS contrast value, the speckle reconstruction, the torchmfbd reconstruction and the NeuralBD reconstruction one can see a clear improvement in the reconstructed observations. The cutouts in the red and green boxes confirm this on different spatial scales. Zooming in to sub-arcsec scales in the green box, NeuralBD shows fine detail structures in the penumbra, which can not be fully resolved by the speckle reconstruction due to noise as well as for the torchmfbd reconstruction. 

Fig.\,\ref{fig:dkist}b) compares the PSD for the single frame of the burst, the speckle reconstruction, the torchmfbd reconstruction and the NeuralBD reconstruction. All three, the speckle, torchmfbd and NeuralBD reconstruction shows more power compared to the single frame of the burst. NeuralBD closely follows the the green line from the single frame of the burst, suggesting that the seeing for this observation was good. Speckle however, shows higher power in the range from 5 arcsec$^{-1}$ to 12 arcsec$^{-1}$ reaching a level similar to that of \mbox{NeuralBD} at around 13 arcsec$^{-1}$. At around 40 arcsec$^{-1}$ the power of speckle reconstruction starts rising, peaking at around 55 arcsec$^{-1}$. This feature is likely caused by noise amplification due to insufficient noise filtering during the speckle reconstruction process. The torchmfbd reconstruction shows higher power starting from around 2 arcsec$^{-1}$, consistently outperforming the original frame, speckle and NeuralBD. The drop in the power at 65 arcsec$^{-1}$ corresponds to the diffraction limit of the telescope.

In Fig.\,\ref{fig:dkist}c) we show five example PSFs estimated from the NeuralBD method. They correspond to the five frames with the best RMS contrast value. The overall structure looks similar to the PSFs estimated for the GREGOR observations in Fig.\,\ref{fig:gregor-psfs}. The PSFs exhibit structured features near their boundaries, suggesting that increasing the PSF size could be beneficial. However, increasing the PSF size would come at the cost of higher memory consumption and longer reconstruction times. Nevertheless, Fig. \ref{fig:dkist} shows that the NeuralBD reconstruction method achieves comparable results to speckle reconstruction while resolving small-scale features slightly more effectively with less noise.

\begin{figure*}[h!tbp]
   \centering
   \includegraphics[scale=0.19]{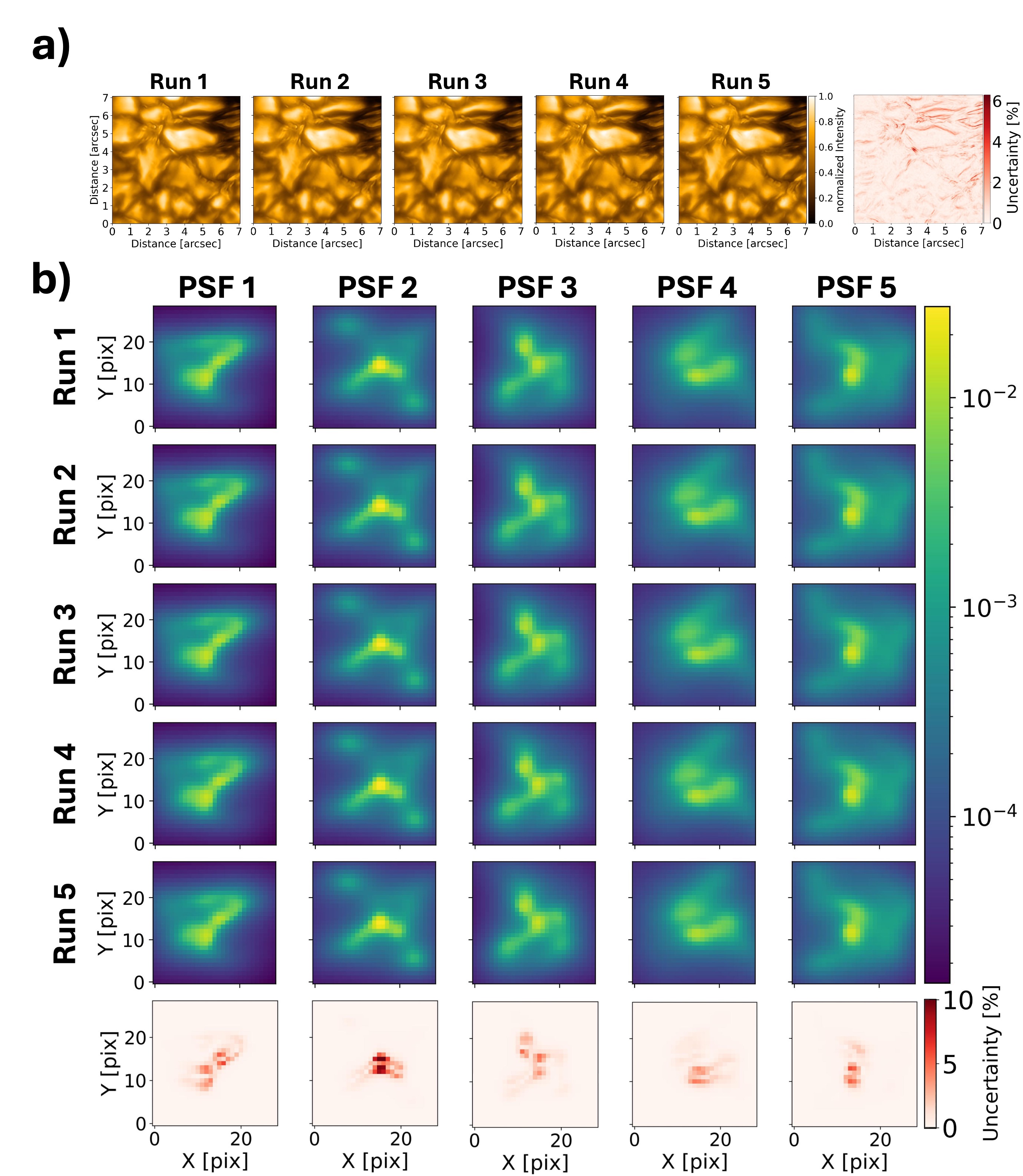}
   \caption{Uncertainty estimation of the NeuralBD model on five individual reconstruction runs using the GREGOR observation. a) shows the NeuralBD reconstruction for the five different runs. The last column corresponds to the standard deviation of the five reconstructions in percent. b) shows five example PSFs for the five different runs. The last row corresponds to the standard deviation evaluated for each of the five PSFs per run.}
\label{fig:uncertainty}
\end{figure*}

\section{Uncertainty estimation}
To assess the uncertainty of our NeuralBD model, we have trained five models with identical configurations using the GREGOR observations as described in Sect.\,\ref{sec:gregor_obs} and Sect.\,\ref{sec:gregor_app}. In Fig.\,\ref{fig:uncertainty}a) we show the NeuralBD reconstruction for the five models. The last column corresponds to the uncertainty map showing the standard deviation (in percent) of the five reconstructions. The reconstructions are visually similar, with consistent large- and small-scale structures across all five runs. The uncertainty map shows only minor variations in the penumbral structures. Figure\,\ref{fig:uncertainty}b) shows five example PSFs from the five different runs. The last row shows the uncertainty map, computed as the standard deviation (in percent) over all 100 PSFs for each of the five runs. The PSFs show similar global structures across all five runs, indicating that the model converges to a similar solution. The uncertainty maps show only small differences between the runs, even though the the overall PSF shapes are nearly identical. These differences are primarily due to sub-pixel spatial shifts of the PSFs, which are also visible in the uncertainty maps. This also explains the small differences in the uncertainty map in the reconstruction in Fig.\,\ref{fig:uncertainty}a).

\section{Discussion}
In this study we developed an image deconvolution method to reconstruct high-resolution ground-based solar observations. The method is based on PINNs and it calculates at the same time the real object intensity distribution and the PSFs responsible for the distortions in the observations. By forward modeling we solve the image equation (Eq.\,\ref{eq:image}) to update the intensity values and the PSF parameters. Using this approach of neural representation can lead to smoother reconstructions with less noise \citep{tancik2020, jarolim2025}. Our method can be directly applied to any high-resolution broadband imaging instruments for ground-based solar observations and has no constraints for telescope aperture and
wavelength. 

We evaluate our NeuralBD method on simulation data to directly compare the  reconstruction performance (see Fig.\,\ref{fig:muram_conv-rec}), and the estimation of the PSFs by our method (Fig.\,\ref{fig:muram_psfs}). Comparing the NeuralBD method to a baseline method and a state-of-the-art multi-frame blind deconvolution method, the NeuralBD reconstruction clearly outperforms both visually (see Fig.\,\ref{fig:muram_rl}) as well as in terms of perceptual and distortion quality metrics as shown in Tab.\,\ref{table:quality-metrics}. Although NeuralBD must infer the PSFs as part of the reconstruction, it achieves higher performance than the baseline method, which benefits from known ground truth PSF. In contrast to the MFBD and MOMFBD approach, NeuralBD does not make any assumptions about the PSFs, such as predefined basis functions or wavefront parameters \citep{vanNoort2005, ramos2018, ramos2023, ramos2024}. This flexibility enables the model to reconstruct the optimal PSFs, and consequently, the most accurate representation of the real object. Compared to previous image reconstruction methods, NeuralBD acts as a solver method and is therefore independent of a high quality data set. Additionally, this makes it independent of the application to any instrument.

Reconstruction of high-resolution observations from the 1.5\,m GREGOR and 4\,m DKIST solar telescopes demonstrated the independent applicability of NeuralBD to different instruments. When compared with state-of-the-art speckle reconstructions and multi-frame blind deconvolution, the NeuralBD results reveal finer details and better resolve small scale features, with reduced noise on the smallest spatial scales due to the smooth image representation provided by the PINN \citep{jarolim2025}. A comparison of the PSFs derived from the GREGOR observations in Fig.\,\ref{fig:gregor-psfs} shows that the NeuralBD method converges towards a unique, physically solution as the PSFs at both wavelengths exhibit similar structural features. This is also confirmed by Fig.\,\ref{fig:uncertainty} where five individual runs show similar performance in the reconstruction as well as close resemblance in the estimated PSFs. Nevertheless, the PSFs derived using NeuralBD differ significantly from the assumed PSFs from MFBD and MOMFBD, which suggests that this additional degree of freedom is essential for accurately modeling the imaging process \citep{ramos2023, ramos2024}. 

By solving the image formation equation (Eq. \ref{eq:image}) directly in the image domain, NeuralBD infers a PSF at each wavelength. This provides several advantages. First, chromatic effects can be handled naturally by estimating wavelength-dependent PSFs. Second, the approach can flexibly accommodate a wide range of image degradations and represent complex PSF structures that are difficult to capture with the parameterization using wavefronts. 
However, working directly in the image domain also has limitations. Unlike wavefront-based parameterizations, the inferred PSFs are not inherently constrained by physical optics. Initializing the PSFs with wavefronts automatically enforce properties such as positivity, normalization and trivial wavelength scaling. Additionally, they often provide a more compact representation of the PSF, if they are dominated by diffraction effects. Furthermore, the large contributions near the PSF boundaries visible in Fig. \ref{fig:dkist} indicate that increasing the PSF size could further improve the reconstructions, which requires more memory and consequently longer reconstruction times.
Wavefront parameterizations can also have inherent limitations. They can struggle to capture effects by phase distortions, such as jittering or other non-diffraction-driven degradations. Moreover, ambiguities arise because different wavefront configurations can produce very similar PSFs, which can complicate the optimization and therfore the reconstruction performance. 
Solving Eq.\,\ref{eq:image} in the image domain and therefore directly estimate the PSFs offers greater flexibility to find an optimal solution, while wavefront-based approaches provide stronger physical constraints and compactness.

The NeuralBD method currently is limited by assuming isoplanatic conditions, restricting the reconstruction to a small FOV. To reconstruct larger FOV one can apply mosaicing of the individual patches. In future studies we will explore pixel-wise spatially variant PSFs to reconstruct the entire FOV of the observations (cf. \citealp{ramos2024}). Currently, the reconstruction process takes around $\sim\,$7 hours on an NVIDIA A100 GPU. In future, we plan to investigate strategies to accelerate this process. This could involve patch-based sampling of the coordinate points, or initializing the network from a pre-trained meta state, in order to reduce the time to reconstruct subsequent observations.

This study demonstrates, that NeuralBD can provide high-resolution solar image reconstruction and exceed the performance to state-of-the-art reconstruction methods in terms of spatial resolution and noise mitigation. Since we have no constraints on telescope aperture, it allows to apply the method to any large aperture ground-based solar telescopes such as the GREGOR telescope, the Swedish solar telescope (SST; \citealp{scharmer2003}), DKIST and the future EST. This study lays the foundation for spatially varying PSFs and has the potential to be extended to other ground-based observations e.g., spectropolarimetric data.

\section*{Data availability}
All codes are publicly available:
\begin{itemize}
    \item Zenodo: \cite{schirninger_neuralbd} and 
    \item GitHub: \url{https://github.com/Schirni/NeuralBD}
\end{itemize}

\begin{acknowledgements}
We thank the referee for very useful comments, which helped to improve the paper. This research has received financial support from the University of Graz EST (European Solar Telescope) program. The research was sponsored by the DynaSun project and has thus received funding under the Horizon Europe programme of the European Union under grant agreement (no. 101131534). Views and opinions expressed are however those of the author(s) only and do not necessarily reflect those of the European Union and therefore the European Union cannot be held responsible for them. RJ was supported by the NASA Jack-Eddy Fellowship. We acknowledge the use of the Vienna Scientific Cluster (VSC) for the computational resources and obtaining the results presented in this paper. This material is based upon work supported by the NSF National Center for Atmospheric Research (NCAR), which is a major facility sponsored by the U.S. National Science Foundation under Cooperative Agreement No. 1852977.
\\
The 1.5-meter GREGOR solar telescope was built by a German consortium under the leadership of the Institute for Solar Physics (KIS) in Freiburg with the Leibniz Institute for Astrophysics Potsdam, the Institute for Astrophysics Göttingen, and the Max Planck Institute for Solar System Research in Göttingen as partners, and with contributions by the Instituto de Astrofísica de Canarias and the Astronomical Institute of the Academy of Sciences of the Czech Republic.
The research reported herein is based in part on data collected with DKIST a facility of the National Science Foundation. DKIST is operated by the National Solar Observatory under a cooperative agreement with the Association of Universities for Research in Astronomy, Inc. DKIST is located on land of spiritual and cultural significance to Native Hawaiian people. The use of this important site to further scientific knowledge is done so with appreciation and respect. We thank Christoph Kuckein for providing us with the speckle reconstructions from the GREGOR telescope and Friedrich Wöger for providing us the DKIST data. \\
This research has made use of AstroPy \citep{astropy2022}, SunPy \citep{sunpy2020} and PyTorch \citep{pytorch2017}. 
\end{acknowledgements}

\bibliographystyle{aa}
\bibliography{bib}
\end{document}